\documentclass{svmult}
\usepackage{mathptmx,helvet,courier,makeidx,graphicx,multicol,footmisc}
\usepackage{psfrag,bbm,amsfonts,amsmath,amssymb,cite}

\newenvironment{proofOF}[2]{\removelastskip\vspace{6pt}\noindent {\it Proof of #1.}~\rm#2}{
\begin{flushright}\qed\end{flushright}}

\newcommand{\Id}{\mathbbm{1}}
\newcommand{\N}{\mathbb{N}}
\newcommand{\Z}{\mathbb{Z}}
\newcommand{\EE}{\mathbb{E}}
\newcommand{\R}{\mathbb{R}}
\newcommand{\Pb}{\mathbb{P}}
\newcommand{\chit}{(t/\chi)^{1/3}}
\newcommand{\chitb}{(t/\chi)^{-1/3}}
\newcommand{\RE}{\mathrm{Re}}
\newcommand{\IM}{\mathrm{Im}}

\newcommand{\Tr}{\mathrm{Tr\,}}
\newcommand{\Ai}{\mathrm{Ai}}
\newcommand{\brem}{\begin{remark}}
\newcommand{\erem}{\end{remark}}
\newcommand{\HH}{\mathcal{H}}
\newcommand{\tHH}{\widetilde{\mathcal{H}}}

\newtheorem{cor}[theorem]{Corollary}

\title*{Convergence of the two-point function of the stationary TASEP}
\author{Jinho Baik\and Patrik L. Ferrari \and Sandrine P\'ech\'e}
\institute{Jinho Baik\at{Department of Mathematics, University of Michigan, Ann Arbor, MI, 48109, USA.} \newline
\email{baik@umich.edu}
\and
Patrik L. Ferrari\at{Institute for Applied Mathematics, Bonn University, Endenicher Allee 60, 53115 Bonn, Germany.} \email{ferrari@uni-bonn.de}
\and
Sandrine P\'ech\'e\at{U.F.R. de Math\'ematiques, Universit\'e Paris Diderot, 75205 Paris, France.} \newline
\email{peche@math.univ-paris-diderot.fr}}

\begin{document}
\maketitle
\sloppy

\abstract{We consider the two-point function of the totally asymmetric simple exclusion process with stationary initial conditions. The two-point function can be expressed as the discrete Laplacian of the variance of the associated height function. The limit of the distribution function of the appropriately scaled height function was obtained previously by Ferrari and Spohn. In this paper we show that the convergence can be improved to the convergence of moments. This implies the convergence of the two-point function  in a weak sense along the near-characteristic direction as time tends to infinity, thereby confirming the conjecture in the paper of Ferrari and Spohn.}

\section{Introduction and result}
The totally asymmetric simple exclusion process (TASEP) is arguably the simplest non-reversible interacting stochastic particle system, and it is also one of the most studied. Particles live on $\Z$ and they satisfy the exclusion constraint: each site can be occupied by at most one particle. Therefore a particle configuration can be denoted by $\eta \in \{0,1\}^\Z$, where $\eta_j=0$ means that site $j$ is empty while $\eta_j=1$ means that the site is occupied. The dynamics of the TASEP is then defined as follows: every particle tries to jump to its right neighbor with rate one. The jumps occurs only if the exclusion constraint is satisfied.

It is known~\cite{Lig76} that the only translation-invariant stationary measures of the TASEP are Bernoulli product measures with parameter $\rho\in [0,1]$, namely,
\begin{equation}
\Pb(\eta_j=1)=\rho \quad \text{for all $j \in \Z$.}
\end{equation}
Here $\rho$ is the average density of particles. The cases $\rho=0$ and $\rho=1$ are trivial and in the following we fix $\rho\in (0,1)$. This system is referred as \emph{stationary TASEP}.

The two-point function is defined as
\begin{equation}
S(j,t):=\EE\left (\eta_j(t)\eta_0(0)\right )-\rho^2.
\end{equation}
Note that this equals the covariance of $\eta_j(t)$ and $\eta_0(0)$.
Hence the two-point function carries the information on how site $j$ at time $t$ is correlated with site $0$ at time~$0$.
It is known that
\begin{equation}
	\sum_{j\in\Z}S(j,t)=\rho(1-\rho)=:\chi
\end{equation}
and also $S(j,t)\geq 0$.
This implies that $\frac1{\chi}S(j, t)$ can be thought of as a probability mass function in $j\in \Z$.
Indeed this equals the probability that a second class particle, which was at site $0$ at time $0$, is at site $j$ at time $t$~\cite{Fer90}.
It is also known that the expectation of $j$ with respect to the probability mass function $\frac1{\chi} S(j,t)$ satisfies
\begin{equation}
\sum_{j\in\Z}j\frac{S(j,t)}{\chi}=(1-2\rho)t,
\end{equation}
and the variance scales as~\cite{BKS85,Sp83}
\begin{equation}
\sum_{j\in\Z}j^2 \frac{S(j,t)}{\chi}-((1-2\rho)t)^2 = O(t^{4/3}).
\end{equation}
as $t\to \infty$.
Therefore, for large time $t$, one expects the scaling form for $S$ as\footnote{The multiplicative factor $\frac{\chi}{4}$ was incorrectly written as $\frac{\chi}2$ in ~\cite{PS01}. This is a typographical error.}
\begin{equation}\label{eq:Sgscsec}
	S(j,t)\simeq \frac{\chi}{4} g_{\rm sc}''\left(\frac{j-(1-2\rho)t}{2\chi^{1/3}t^{2/3}}\right) \frac1{2\chi^{1/3}t^{2/3}}
\end{equation}
for some non-random function $g_{\rm sc}$.
The precise expression of $g_{\rm sc}$ was first conjectured in~\cite{PS02b} based on the work~\cite{BR00}:
\begin{equation}\label{eq:gscFwse}
	g_{\rm sc}(w)=\int_{\R} s^2 dF_w(s)
\end{equation}
where $F_w(s)$ is the distribution function defined~\eqref{eqFw2} below.

In order to understand the presence of the second derivative in~\eqref{eq:Sgscsec} and the second moment formula~\eqref{eq:gscFwse}, we recall that TASEP can also be seen as a stochastic growth interface model, whose discrete gradient of the height equals $1-2\eta$. The dynamical rule is that when a particle jumps to the right, a valley $\diagdown\diagup$ changes to a mountain $\diagup\diagdown$.
More precisely, let $N_t(j)$ denote the number of particles which have jumped  from site $j$ to $j+1$ during the time interval $[0,t]$, and define the height function
\begin{equation}
h_t(j)= \begin{cases}
  2N_t(0) +\sum_{i=1}^j (1-2\eta_i(t))& \textrm{for }j\geq 1,\\
2N_t(0) & \textrm{for }j=0,\\
2N_t(0) -\sum_{i=j+1}^0 (1-2\eta_i(t))& \textrm{for }j\leq -1.
 \end{cases}
\end{equation}
Then initially $h_0(0)=0$ and \mbox{$h_0(j)-h_0(j-1)=1-2\eta_j(0)$},
and at the instance a particle jumps from site $j$ to $j+1$, the height function at position $j$ increases by two.
Note that $h_t(j)-h_0(j)=2N_t(j)$.
It was shown in~\cite{PS01} that the two-point function can be expressed as
\begin{equation}\label{eqS}
S(j,t)=\tfrac18 \big(\Delta{\rm Var}(h_t(\cdot))\big)(j)
\end{equation}
with $\Delta$ being the discrete Laplacian, $(\Delta f)(j)= f(j-1)-2f(j)+f(j+1)$.
Since it is known that $F_w(s)$ has mean $0$~\cite{BR00},
this explains the presence of the second derivative in the conjectured formula~\eqref{eq:Sgscsec} and the second moment formula~\eqref{eq:gscFwse}.

Define the probability distribution functions of the location-rescaled height function,
\begin{multline}\label{eqFwt}
F_w(s,t):= \Pb\Big(h_{t}([(1-2\rho)t+2w \chi^{1/3}t^{2/3}]) \\
\geq (1-2\chi)t +2w(1-2\rho)\chi^{1/3}t^{2/3}-2s\chi^{2/3}t^{1/3}\Big).
\end{multline}
The function $F_w$ in~\eqref{eq:gscFwse} (which is defined in~\eqref{eqFw2} below) was conjectured in \cite{PS02b} to be the limit
\begin{equation}\label{eqFw}
	\lim_{t\to\infty} F_w(s,t)= F_w(s).
\end{equation}
The convergence~\eqref{eqFw} for each $s$ was later proved in~\cite{FS05a}.
This strongly indicates the validity of~\eqref{eq:Sgscsec}. A missing part in concluding~\eqref{eq:Sgscsec} is the convergence of the moments of $F_w(s,t)$ which is a stronger statement than~\eqref{eqFw}. Our main result is that the moments indeed converge.

\begin{theorem}\label{Ourthm}
For all $\ell\in \N$,
\begin{equation}\label{eqToBeProven}
	\lim_{t\to\infty} \int_\R s^\ell dF_w(s,t)=\int_\R s^\ell dF_w(s)
\end{equation}
uniformly for $w$ in a compact subset of $\R$.
\end{theorem}
As a consequence we obtain the convergence of the two-point function is a weak sense.

\begin{cor}\label{corollario}
We have, with $\chi:=\rho(1-\rho)$,
\begin{equation}\label{eq:Sconv}
\lim_{t\to\infty}2\chi^{1/3}t^{2/3}S\left([(1-2\rho)t+2w\chi^{1/3}t^{2/3}],t\right)=\frac{\chi}{4} g_{\rm sc}''(w)
\end{equation}
if integrated over smooth functions in $w$ with compact support.
\end{cor}

The proof of this corollary is given in Section~\ref{sectCor}. An improvement of the analysis in this paper can yield the convergence in the point-wise sense in~\eqref{eq:Sconv}. However, we do not consider this direction in this paper.

\medskip

For completeness, let us state a formula of the limiting distribution function $F_w(s)$ explicitly.
Let $P_u$  be the orthogonal projector on the interval $[u, +\infty)$.
Set
\begin{equation}
\begin{aligned}
K_{\Ai,s}(x,y)&:=\int_{\R_+}\Ai(x+s+\lambda)\Ai(y+s+\lambda) d\lambda,\\
F_{\rm GUE}(s)&:=\det(\Id-P_0K_{\Ai,s}P_0).
\end{aligned}
\end{equation}
$F_{\rm GUE}$ is the GUE Tracy-Widom distribution function~\cite{TW94}. We also define the function
\begin{equation}
	g(s,w):=e^{-\frac{w^3}{3}}\bigg( \int_{\R^2_-} e^{w(x+y)}\Ai(x+y+s)dx\, dy
+ \int_{\R^2_+} \widehat \Psi_{w,s}(x)\rho_s(x,y) \widehat\Phi_{w,s}(y) dx\, dy\bigg),
\end{equation}
where
\begin{equation}
\widehat \Phi_{w,s}(x):=\int_{\R_-}e^{wz+ws}K_{\Ai,s}(z,x)dz, \quad \widehat \Psi_{w,s}(x):=\int_{\R_-}e^{wz}\Ai(x+z+s)dz,
\end{equation}
and $\rho_s(x,y):=(\Id-P_0K_{\Ai,s}P_0)^{-1}(x,y)$.
Now
\begin{equation}\label{eqFw2}	
	F_w(s):=\frac{\partial}{\partial s}\left(F_{\rm GUE}(s+w^2)g(s+w^2,w)\right).
\end{equation}
There is an alternative formula expressed in terms the Lax pair equations of the Painlev\'e II equation obtained in \cite{BR00}. But we will only use the formula~\eqref{eqFw2} in this paper.
One can also consider the joint distributions for different values of $w$ and a formula can be found in~\cite{BFP09}.

\subsubsection*{Acknowledgments}
We would like to thank Ivan Corwin and Eric Cator for a communication which helped us simplify the proof of  Proposition~\ref{thm:tail2}.
The work of Jinho Baik was supported in part by NSF grants DMS1068646. Patrik Ferrari was supported by the German Research Foundation via the SFB611--A12 project.

\section{Setting and strategy of the proof}
The height function $h_t(j)$ associated to a TASEP with any initial condition can be
related to the last passage time of a directed last passage percolation (DLPP) model.
Over the last decade or so, the so-called solvable, or determinantal DLPP models \cite{Jo00b, Ok02, BP08}
were studied extensively.
These are the models for which the probability distribution of the last passage time can be expressed explicitly in terms of Fredholm determinants.
The DLPP model corresponding to the stationary TASEP is not one of solvable models but can be related to one after suitable analytic continuation of the parameters.
This yields the following formula of $F_w(s,t)$.

Fix $w\in\R$. Let us set\footnote{To be precise, we need to take the integer parts of the formulas since $m$ and $d$ need to be integers. Since the error between the formula above and the integer parts is $O(1)$, this does not result in any significant changes in the estimates and hence for convenience we define $m$ and $d$ as in~\eqref{eq:defmd} without restricting them to be integers in this paper. However, we remark that if we restrict $m$ and $d$ to be integers, one occasionally need to be careful in the precise formulation of the estimates and the exposition becomes more involved. We do not discuss these subtleties in this paper.\label{footn}}
(recall that $\chi=\rho(1-\rho)$)
\begin{equation}\label{eq:defmd}
	2m=(1-2\chi)t+2w(1-2\rho)\chi^{1/3}t^{2/3},\quad 2d=(1-2\rho)t+2w\chi^{1/3}t^{2/3},
\end{equation}
and define the functions~\footnote{For any set of points $S$, the notation $\oint_{\Gamma_{S}}  f(z) \, dz$ denotes the integral over a simple closed contour which encloses the points $S$ but excludes any other poles of the function $f$. The contour is oriented counter-clockwise.}
\begin{equation}
\begin{aligned}
L(x,y)&=\frac{-e^{a(x-y)}}{2\pi\I}\oint_{\Gamma_{1-\rho}}e^{-z(x-y)}\frac{(z+\rho)^{m-d}}{(1-\rho-z)^{m+d}} dz \quad \textrm{ for }x>y, \\
R(x,y)&=\frac{e^{a(x-y)}}{2\pi\I}\oint_{\Gamma_{-\rho}}e^{z(y-x)}\frac{(1-\rho-z)^{m+d}}{(\rho + z)^{m-d}} dz \quad \textrm{ for }x<y,
\end{aligned}
\end{equation}
with
\begin{equation}\label{eq:adef}
	a:=\frac12-\rho.
\end{equation}
We define the kernel
\begin{equation}
	K_{m,d}(x,y)=\int_{\R_-}L(x,z)R(z,y)dz,
\end{equation}
and the distribution function
\begin{equation}\label{eq:Ff}
	F(u):= \det(\Id-P_uK_{m,d}P_u).
\end{equation}
Finally, we set
\begin{equation}\label{eq:Gf}
	G_0(u)=g_1(u)+g_2(u)+g_3(u),
\end{equation}
where
\begin{equation}\label{eq:defg123}
\begin{aligned}
g_1(u)&=u+\frac{2ad-m}{1/4-a^2},\\
g_2(u)&=\langle \psi_a, P_u K_{m,d}\psi_{-a}\rangle,  \\
g_3(u)&=\langle K_{m,d}^*(\Id-P_u)\psi_a, P_u(\Id-P_uK_{m,d}P_u)^{-1}P_u(\Id-K_{m,d})\psi_{-a}\rangle,
\end{aligned}
\end{equation}
with $\psi_a(x)=e^{-ax}$. Then it was shown in~\cite{FS05a} that\footnote{The formula (\ref{eq:24})
is the formula (4.10) of~\cite{FS05a} when $b=-a$ if we take into account~\eqref{eq:25} .
See (5.21) of~\cite{FS05a} for the formula of the function $G_0(u)=G^{a,-a}(u)$.}
\begin{equation}\label{eq:24}
	F_w(s,t)=\frac{1}{\chit}\frac{d}{ds}\left ( F(u(s,t))G_0(u(s,t))\right)
\end{equation}
where
\begin{equation}\label{eq:25}
	u=u(s,t):=t+s\chi^{-1/3}t^{1/3}.
\end{equation}

Set
\begin{equation}
\widehat G_0(s,t):=G_0(u(s,t)), \qquad \widehat F(s,t):=F(u(s,t)).
\end{equation}
The main technical part of this paper is the following estimates:\footnote{The exponents of the bounds are not optimal. The bound in~\eqref{eq:up} and~\eqref{eq:low} can be improved to $Ce^{-c|s|^{3/2}}$ and $Ce^{-c|s|^3}$ respectively. The improved bound for~\eqref{eq:up} can be achieved if we keep track of a slightly better estimate in the analysis presented in this paper. On the other hand, in order to improve the bound~\eqref{eq:low}, we need a different approach such as Riemann-Hilbert analysis as in~\cite{BDJ99, BDMMZ01}.}

\begin{proposition}[Uniform upper tail estimates]\label{thm:tail1}
There exist positive constants $s_0$, $t_0$, $c$ and $C$ such that
\begin{equation}\label{eq:up}
	\left| s-\widehat F(s,t) \widehat G_0(s,t)/\chit \right| \leq Ce^{-c|s|}, \qquad s\ge s_0, \quad t\ge t_0.
\end{equation}
The bound holds uniformly for $w$ in a compact subset of $\R$.
\end{proposition}

\begin{proposition}[Uniform lower tail estimates]\label{thm:tail2}
There exist $s_0$, $t_0$, $c$ and $C$ such that
\begin{equation}\label{eq:low}
	\left| F_w(s,t) \right| \leq C e^{-c|s|^{3/2}}, \qquad s\le -s_0,\quad  t\ge t_0.
\end{equation}
The bound holds uniformly for $w$ in a compact subset of $\R$.
\end{proposition}

Theorem~\ref{Ourthm} now follows.

\begin{proofOF}{Theorem~\ref{Ourthm}}
We only consider $\ell\ge 2$. The case $\ell=1$ follows easily.
We first write the integral on the left-hand-side of~\eqref{eqToBeProven} as the sum of the integral over $\R_+$ and the integral over $\R_-$.
For the integral over $\R_+$, integrating by parts twice and using the fact that $F_w(\cdot, t)$ is a cumulative distribution function,
\begin{equation}\label{eq:mom1}
\begin{aligned}
	\int_{\R_+} s^\ell dF_w(s,t) =& -\ell(\ell-1) \int_{\R_+}s^{\ell-2}\Big(s-\widehat F(s,t) \frac{\widehat G_0(s,t)}{\chit}\Big)ds
\end{aligned}
\end{equation}
for $\ell\ge 2$.
It was in~\cite{FS05a} that in addition to~\eqref{eqFw} we also have limit \mbox{$\widehat F(s,t) \frac{\widehat G_0(s,t)}{\chit}\to F_{\rm GUE}(s+w^2)g(s+w^2,w)$} for each $s$ as $t\to\infty$.
Thus due to Proposition~\ref{thm:tail1} the Lebesgue dominated convergence theorem can be applied and we find that~\eqref{eq:mom1} converges to
\begin{equation}\label{eq:mom2}
\begin{aligned}
	-\ell(\ell-1) \int_{\R_+}s^{\ell-2}\Big(s-F_{\rm GUE}(s+w^2)g(s+w^2,w)\Big)ds.
\end{aligned}
\end{equation}

On the other hand, integrating by parts once,
\begin{equation}\label{eq:mom3}
\begin{aligned}
	\int_{\R_-} s^\ell dF_w(s,t) =&-\ell \int_{\R_-} s^{\ell-1}F_w(s,t) ds.
\end{aligned}
\end{equation}
Thus again, using the Lebesgue dominated convergence theorem can be applied due to Proposition~\ref{thm:tail2} and
from~\eqref{eqFw} we find that~\eqref{eq:mom3} converges to
\begin{equation}\label{eq:mom4}
\begin{aligned}
	-\ell \int_{\R_-} s^{\ell-1}F_w(s) ds.
\end{aligned}
\end{equation}

Integrating~\eqref{eq:mom2} and~\eqref{eq:mom4} by parts backwards and using the fact that $F_w$ is a cumulative distribution function, we find that the sum of these two integrals is the right-hand-side of~\eqref{eqToBeProven}.
\end{proofOF}

The estimate~\eqref{eq:up} for the upper tail is obtained by analyzing the formulas~\eqref{eq:Ff} and~\eqref{eq:Gf} asymptotically using the saddle-point analysis. This asymptotic analysis is very close to that of many previous papers, for example~\cite{Jo00b, FS05a, BBP06}. We use some of the results directly or improve upon them. See Section~\ref{sec:up}.

For the estimate~\eqref{eq:low} on the lower tail, we note the following.
Consider the TASEP with step-initial condition i.e. $\eta_j(0)=1$ for $j\le 0$ and $\eta_j(0)=0$ for $j\ge 1$.
Then the associated height function $h_t^{\rm step}(j)$ satisfies $h_0^{\rm step}(j)=|j|$. This means that initially $h_0$ is bounded above by $h_0^{\rm step}$. Since the initial condition of the stationary TASEP is independent of the dynamics,
we find that $h_t$ is stochastically bounded above\footnote{This can also be seen easily from the corresponding directed last passage percolation (DLPP) models. The DLPP model for the stationary TASEP
is the DLPP model for the TASEP with the step initial condition plus an extra row and an extra column with non-zero weights.} by $h_t^{\rm step}$.
Hence\footnote{We would like to thank Ivan Corwin and Eric Cator for communicating this observation  with us.
This observation simplified the proof of the lower tail estimate which we originally obtained by estimating $F_w(s,t)$ directly.}
\begin{equation}\label{eq8-1}
	\Pb( h_t(j) \ge u ) \leq \Pb( h_t^{\rm step}(j)\ge u ).
\end{equation}
But $\Pb( h_t^{\rm step}(j)\ge u )$ is known to be precisely $F(u)$ of~\eqref{eq:Ff}~\cite{Jo00b}.
Therefore we have
\begin{equation}\label{eq8}
	F_w(s,t)\leq \widehat F(s,t)=\det(1-P_u K_{m,d} P_u).
\end{equation}
Thus the estimate~\eqref{eq:low} follows if we show that $\widehat F(s,t)$ is bounded above by $Ce^{-c|s|^{3/2}}$ for negative large enough $s$.
This in turn follows if we show the same bound for the Fredholm determinant~\eqref{eq:Ff}.
For this purpose we  follow the idea of Widom~\cite{Wid02} which seems not as well-known as it should be.
See Section~\ref{sec:low}.

\section{Proof of Proposition~\ref{thm:tail1}: upper tail}\label{sec:up}

The proposition follows from~\eqref{eq:32},~\eqref{majohatFs>0} and ~\eqref{eq21-1} in the below.

\subsection{Asymptotics for $\widehat F$}\label{sec:ashatFup}

The  function $F(u)=\det(1-P_{u}K_{m,d}P_{u})$ is the distribution function of the last passage time of the directed last passage model with i.i.d. exponential random variables.
It is well-known~\cite{Jo00b} that this also equals the distribution function of the largest eigenvalue of the Laguerre unitary ensemble (LUE)
which is defined as $M_{m,d}=\frac{1}{m-d}XX^*$ where $X$ is a $(m-d)\times (m+d)$ random matrix with i.i.d. standard complex Gaussian entries.
This equality can also be seen explicitly in Appendix C of~\cite{FS05a}
where $K_{m,d}$ was shown to be same as the correlation kernel of the LUE up to a  conjugation by a multiplication.
The asymptotics of LUE and $\widehat F(s,t)=F(u(s,t))$ were considered in several papers, especially in~\cite{Jo00b, FS05a, BBP06}. We have:

\begin{lemma}
Fix $s_0\in \R$. Then
\begin{equation}
	\lim_{t\to \infty}\widehat F(s,t)
	=F_{\rm GUE}(s+w^2)
\end{equation}
uniformly for $s\in [s_0, \infty)$ and $w$ in a compact subset of $\R$.
Furthermore, for given $s_0\in \R$ and $t_0>0$, there exist positive constants $C$ and $c$ such that
\begin{equation}\label{majohatFs>0}
	|1-\widehat F(s,t)|\leq Ce^{-c s}
\end{equation}
for $s\ge s_0$ and $t\ge t_0$.
\end{lemma}

The bound (\ref{majohatFs>0}) can be found in, for example, Section 3.1 of~\cite{BBP06}.\footnote{The exponent of the upper bound is not optimal: the optimal exponent is $e^{-c|s|^{3/2}}$. But we do not consider such an issue in this paper.}

\subsection{Evaluation of $g_1$}\label{sec:evg1}

A direct computation using~\eqref{eq:defmd},~\eqref{eq:adef}, and~\eqref{eq:25} shows that\footnote{The formula becomes $s\chit+O(1)$ where $O(1)$ is independent of $s$ if we take the integer parts in the definition of $m$ and $d$ in~\eqref{eq:defmd}.
This is an example of the subtleties mentioned in the Footnote~\ref{footn}. This results in the additional term $O(t^{-1/3})$ in~\eqref{eq:32}. Since this is not a function in $s$, we cannot obtain the bound (C1). However, this issue can be fixed by shifting $s$ to $s-O(1)/\chit$. In other words, the centering and scaling $u=t+s\chit$ needs to be changed slightly to reflect the difference of the formula of~\eqref{eq:defmd} and their integer counter-parts.}
\begin{equation}\label{g1maj}
g_1(u)=u+\frac{2ad-m}{1/4-a^2}=s\chit.
\end{equation}
This implies that
\begin{equation}\label{eq:32}
	s- \frac{\widehat F(s,t) \widehat G_0(s,t)}{\chit}= - \widehat F(s,t) \frac{g_2(u)+g_3(u)}{\chit} + s (1- \widehat F(s,t) )
\end{equation}
The term $1- \widehat F(s,t)$ can be estimated using~\eqref{majohatFs>0} and $\widehat F(s,t)$ is bounded by $1$ since it is a distribution function.  We now show that
$g_2(u)/\chit$ and $g_3(u)/\chit$ are uniformly (in $t$) bounded by exponentially decaying functions in $s$.

In the rest of this section, we only consider the case when $w> 0$.
If $w<0$, we need to start with a different decomposition of $G_0(u)$ ((5.22) instead of (5.21) of~\cite{FS05a}).
After this change, the analysis is completely analogous.
For the case when $w=0$, we can proceed as in the case when $w>0$ but with a yet slight modification: see (6.31)-(6.34) of~\cite{FS05a}.
We skip the detail when $w<0$ and $w=0$, and assume from now on that $w>0$.

\subsection{Estimations on $g_2$ and $g_3$}

Recall the definition~\eqref{eq:defg123} of $g_2(u)$.
It is a direct calculation to show that (see  (3.15) of~\cite{FS05a})
\begin{equation}\label{eq:RPhi}
	\int_x^\infty R(x,y) \psi_{-a}(y) dy= Z(\rho) \psi_{-a}(x), \qquad Z(\rho):=\frac{(1-\rho)^{m+d}}{\rho ^{m-d}},
\end{equation}
for $x\in\R$, for $a\in (-1/2, 1/2)$. Using this, $g_2(u)= \langle \psi_a, P_u L(\Id-P_0) \psi_{-a} \rangle$.
Inserting the formula  $\psi_a$ and $L(x,y)$, we obtain
\begin{equation}\label{limHN-1}
	g_2(u) = \int_{\R_+^2} \HH_t(x+y) dx\, dy,
\end{equation}
where
\begin{equation}\label{def:HN}
	\HH_t(x):= \frac{-Z(\rho)}{2\pi\I}\oint_{\Gamma_{1- \rho}}e^{-z(u+x)}\frac{(z+\rho)^{m-d}}{(1-\rho-z)^{m+d}} dz.
\end{equation}
Thus (see (6.19) of~\cite{FS05a})
\begin{equation}\label{limHN}
	\chitb g_2(u) = \int_{\R_+^2}H_t(x+y) dx\, dy,
	\quad H_t(y):= \chit \HH_t(y\chit).
\end{equation}

Similarly, recall the definition~\eqref{eq:defg123} of $g_3(u)$.
Using~\eqref{eq:RPhi}, an argument similar to that for~\eqref{limHN-1} implies that
\begin{equation}\label{eq:q1}
	(\Id- K_{m,d})\psi_{-a}(x)= e^{ax} \bigg[ 1- \int_{\R_+} \HH_t(-u+x+y) dy \bigg].
\end{equation}
We also note that, similar to~\eqref{eq:RPhi}, we have (see (3.15) of~\cite{FS05a})
\begin{equation}\label{eq:LPhi}
	\int_x^\infty \psi_{a}(y)L(y,x) dy= \frac1{Z(\rho)} \psi_{a}(x)
\end{equation}
for $x\in \R$, for $a\in (-1/2, 1/2)$. Using this, we find that
\begin{multline}\label{eq:q2}
	K_{m,d}^* (1-P_u)\psi_{a}(x) \\
	 = e^{-ax} \bigg[ \int_{\R_+} \tHH_t(-u+x+y) dy
	- \int_{\R_+^2} \tHH_t(-u+x+y)\HH_t(z+y) dz dy \bigg]
\end{multline}
where
\begin{equation}\label{def:q3}
	\tHH_t(x):= \frac{1}{2\pi\I \,Z(\rho)}\oint_{\Gamma_{- \rho}}e^{z(u+x)}\frac{(1-\rho-z)^{m+d}} {(z+\rho)^{m-d}}dz.
\end{equation}
This implies that we can express (see (6.26)-(6.28) in~\cite{FS05a})
\begin{equation}
\chitb g_3(u)=\langle \Phi_t, A_t \Psi_t\rangle \label{defg3}
\end{equation}
where
\begin{equation}
\begin{aligned}
\label{def: PhiNPsiN}
\Phi_t(\xi)&=e^{w\xi}\Big[\int_{\R_+}\widetilde H_t(y+\xi)dy-\int_{\R_+^2} H_t(x+y)\widetilde H_t(y+\xi)dx\,dy\Big],\\
\Psi_t(\xi)&=e^{-w\xi}\Big[1-\int_{\R_+}H_t(y+\xi)dy\Big],
\end{aligned}
\end{equation}
with $H_t(y)= \chit \tHH(y \chit)$ and $\widetilde H_t(y)= \chit \tHH_t(y \chit)$,
and the operator $A_t$ is defined by $A_t=P_0(\Id-K_t)^{-1}P_0$ where the kernel of $K_t$ is
\begin{equation}
	K_t(\xi_1,\xi_2)=e^{w(\xi_2-\xi_1)}\int_{\R_+}H_t(x+\xi_1) \widetilde H_t(x+\xi_2)dx, \qquad \xi_1, \xi_2\ge 0,
\end{equation}
and $K_t(\xi_1,\xi_2)=0$ otherwise.

We obtain the following estimates for $g_2$ and $g_3$.

\begin{lemma} \label{lem: g2s>0}
There are positive constants $c$ and $C$ such that
\begin{equation}\label{eq21-1}
	\left|\chitb g_2(u) \right|\leq Ce^{-cs}, \qquad |\chitb g_3(u)|\leq Ce^{-cs}
\end{equation}
for all $s\geq 0$ and $t\ge 0$.
\end{lemma}

\begin{proofOF}{Lemma~\ref{lem: g2s>0}}
Note from the formula~\eqref{def:HN} that $\HH_t(x)= \HH_t(x; u)$ is a function of $x+u$ .
Hence $H_t(y)=H_t(y;s)$ is a function of $y+s$. Thus, $H_t(y;s)=H_t(y+s;0)$. The same holds for $\widetilde H_t(y)= \widetilde H_t(y;s)$.

Basic bounds for the functions $H_t(y)$ and $\widetilde H_t(y)$ were obtained in (6.15) of~\cite{FS05a}: for any $\beta >0$ there exist positive constants $C_{\beta}$ and $C'_{\beta}$ such that
\begin{equation}\label{eq36}
	|H_t(y;s)|\leq C_{\beta}e^{-\beta y} \quad \textrm{and} \quad |\widetilde H_t(y;s)|\leq C'_{\beta}e^{-\beta y}
\end{equation}
uniformly for $t\ge 0$, $y\ge 0$, and $s\ge 0$.
In particular, the bound holds for $H_t(y;0)$ and $\widetilde H_t(y;0)$ when $s=0$, for $t\ge 0$ and $y\ge 0$.
Thus using $H_t(y;s)=H_t(y+s;0)$ and inserting $y+s$ in place of $y$ in~\eqref{eq36}, we find that:
for any $\beta>0$ there are positive constants $C_{\beta}$ and $C'_{\beta}$ such that
\begin{equation}\label{eq36-1}
	|H_t(y;s)|\leq C_\beta e^{-\beta(s+y)} \quad \textrm{and} \quad |\widetilde H_t(y;s)|\leq C'_\beta e^{-\beta(s+y)}
\end{equation}
uniformly in $t\ge 0$, $y\ge 0$ and $s\ge 0$.

The bound for $\chitb g_2(u)$ follows from~\eqref{limHN} and~\eqref{eq36-1}.

We now estimate $|\chitb g_3(u)|$.
Choosing $\beta >|w|$,~\eqref{eq36-1} implies that \mbox{$|\Phi_t(\xi)|\leq C e^{-\beta s} e^{-(\beta-w)\xi}$} for a positive constant $C$.
Thus,
\begin{equation}\label{eq37}
	|| \Phi_t||_{L^2(\R_+)}\leq C' e^{-\beta s},
\end{equation}
for a constant $C'$ uniformly in $t\ge 0$ and $s\ge 0$.
On the other hand,~\eqref{eq36-1} implies that $|\Psi_t(\xi) e^{w\xi}|$ is bounded by a constant.
Since we assume $w> 0$ (see Section~\ref{sec:evg1}), we find that
$|| \Psi_t||^2_{L^2(\R_+)}$ is uniformly bounded in $t\ge 0$ and $s\ge 0$.
Finally, using the inequality
\begin{equation}\label{eq:Ates}
	||A_t||\leq ||(\Id-K_{\Ai, w^2+s})^{-1}||+ ||(\Id-K_{\Ai, w^2+s})^{-1}-(\Id-K_t)^{-1} ||
\end{equation}
where $K_{\Ai, w^2+s}$ is the Airy kernel restricted on $(w^2+s, \infty)$, and the fact (see (6.36) of~\cite{FS05a}) that
$||(\Id-K_{\Ai})^{-1}-(\Id-K_t)^{-1} ||\to 0$ as $t\to \infty$ imply that $||A_t||$ is uniformly bounded in $t\ge 0$ and $s\ge 0$.
Therefore, the bound for $\chitb g_3(u)$ follows from $|\langle \Phi_t, A_t \Psi_t\rangle |\leq ||\Phi_t||\, ||A_t ||\, || \Psi_t||$.
\end{proofOF}

\section{Proof of Proposition~\ref{thm:tail2}: lower tail}\label{sec:low}

Recall from Section~\ref{sec:ashatFup} that $K_{m, d}$ is a similarity transform of the correlation kernel of the LUE $M_{m,d}$. Since the correlation kernel of the LUE is a positive projection,
all the eigenvalues, which we denote by $\mu_j, j= 0,1,2,\cdots,$ of $P_uK_{md}P_u$ are real and $\mu_j\in [0,1]$.
It was shown in Appendix B.3 of~\cite{FS05a} that $\mu_j \in [0,1)$ if $u>0$.
From this we find that $\det(1-P_u K_{m,d} P_u)=\prod_{j\geq 0} (1-\mu_j)\le \prod_{j\geq 0} e^{-\mu_j}= e^{-\Tr (P_u K_{m,d} P_u)}$.
Therefore,
\begin{equation}\label{eq:wFupesT}
	\widehat F(s,t) \leq \exp(-\Tr(P_u K_{m,d} P_u)).
\end{equation}
This  trick is due to Widom~\cite{Wid02}.

The trace has the following lower bound:

\begin{proposition}\label{TheoBoundTrace}
There exist positive constants $t_0$, $s_0$, $c$ such that
\begin{equation}\label{eq:Trestcs32}
	\Tr(P_u K_{m,d} P_u) \geq c |s|^{3/2}
\end{equation}
for all $s\le -s_0$ and  $t\geq t_0$.
\end{proposition}

The same estimate was obtained in the context of random permutations and an oriented digital boiling model by Widom~\cite{Wid02}. We follow the paper~\cite{Wid02} to prove the Proposition, and as such we only sketch the main ideas and do not provide all the details of the proof.
Once this proposition is proved, then Proposition~\ref{thm:tail2} follows from~\eqref{eq8} and~\eqref{eq:wFupesT}.

\begin{proofOF}{Proposition~\ref{TheoBoundTrace}}
Since the operator $K_t$ is trace class with continuous kernel, we have
\begin{equation}\label{trkn}
\begin{aligned}
\Tr(K_t)=\int_{\mathbb{R_+}} K_t(x,x)dx\,
&=\frac{-1}{(2\pi\I)^2}\oint_{\Gamma_1}\oint_{\Gamma_0} \frac{e^{wu}}{e^{zu}}\frac{(1-w)^{m+d}}{(1-z)^{m+d}}\frac{z^{m-d}}{w^{m-d}}\frac{dw\, dz}{(w-z)^2}\\
&=\frac{-1}{(2\pi\I)^2}\oint_{\Gamma_1}\oint_{\Gamma_0} \frac{e^{M F_{u}(w)}}{e^{M F_{u}(z)}} \frac{dw\, dz}{(w-z)^2}
\end{aligned}
\end{equation}
where
\begin{equation}\label{eq:Fud}
	F_{u}(z):= u'z-\ln z +\gamma \ln (1-z), \qquad u':= \frac{u}{M}.
\end{equation}
Here $M:=m-d$ and \mbox{$\gamma:=\frac{m+d}{m-d}=(\frac{1-\rho}{\rho})^2 + O(t^{-1/3})$}.
Note that $u'= \frac1{\rho^2} + O(t^{-1/3})$ if $s$ is in a bounded set, and $u'\ge \frac1{\rho^2}$ for all $s\le 0$.
We analyze  (\ref{trkn}) asymptotically using the saddle-point analysis.
Note the presence of the singularity $\frac{1}{(w-z)^2}$ in the integrand.

We first consider the case where
\begin{equation}\label{eq:condup1}
	(1-\sqrt{\gamma})^2+\epsilon\le u'<(1+\sqrt \gamma)^2-s_0 t^{-2/3},
\end{equation}
for some $\epsilon>0$ (small, but fixed) and $s_0\gg 1$ also fixed. The critical points are $F_{u}$ are
\begin{equation}\label{critpoints}
	z_c^{\pm}(u')=\frac{u'+1-\gamma}{2u'}\pm \frac{1}{2u'}\sqrt{(u'-(1+\sqrt{\gamma})^2)(u'-(1-\sqrt{\gamma})^2)}.
\end{equation}
The two critical points are non-real and  $|z_c^{\pm}(u')|=\frac1{\sqrt{u'}} \le \rho<1$.
Consider the following two contours:
\begin{equation}\label{eq63}
	w = |z_c^+| e^{\I \theta},\qquad  0\leq \theta <2\pi, \end{equation}
and
\begin{equation}\label{eq63-2}
	z = 1+|z_c^+-1|e^{\I \theta}, \qquad 0\leq \theta <2\pi,
\end{equation}
respectively. Then
\begin{equation}
\begin{aligned}
\RE \left (\frac{d}{d\theta}F_{u}(w)\right)&=-\IM(w) \left (u'-\frac{\gamma}{|w-1|^2}\right )
=-\IM (w) \left (\frac{\gamma}{|z_c^+-1|^2}-\frac{\gamma}{|w-1|^2}\right ), \\
\RE \left (\frac{d}{d\theta}F_{u}(z)\right )&=-\IM(z) \left (u'-\frac{\gamma}{|z|^2}\right )=-\IM (z) \left (\frac{\gamma}{|z_c^+|^2}-\frac{\gamma}{|z|^2}\right ).
\end{aligned}
\end{equation}
Thus along these contours $\RE \left( F_{u}\right )$ achieves its relative maximum (resp. minimum) at $z_c^{\pm}$. Hence these paths are of steep-ascent and steep-descent for $F_u$.
We chose to work with these explicit contours instead of the contours of steepest-ascent and steepest-descent for convenience.
Due to this reason, we need to modify the contours locally near the critical points if $u'$ is close to $(1+\sqrt{\gamma})^2$. Namely, in this case, the contours above become almost tangential and are almost parallel to the direction along which $\RE (F_u)$ is constant.
Then we cannot apply the saddle-point method. In this case, we simply modify the contours locally
near the critical points so that it passes through the critical points along the steepest descent direction as pictured in Figure~\ref{FigSmallS} for the $z$-contour. A similar modification is needed for the $w$-contour.
This small modification does not yield any significant changes in the estimation. For the convenience of presentation, we work with the above explicit contours and skip the details on how the formulas changes after the modifications. The same procedure was also explained in Section~6.2 of~\cite{BF08} for the similar estimations.
\begin{figure}[t]
\begin{center}
\psfrag{0}{$0$}
\psfrag{1}{$1$}
\psfrag{zc}{$z_c^+$}
\psfrag{G}{$\Gamma$}
\psfrag{z}[r]{$\sqrt{1/u'}$}
\psfrag{1-z}{$\sqrt{\gamma/u'}$}
\includegraphics[height=4.5cm]{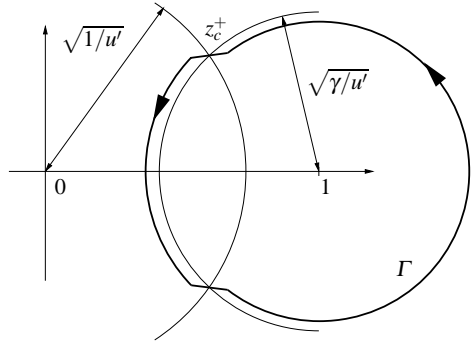}
\caption{The bold path $\Gamma$ is the deformation of $\Gamma_1$ that locally follows the steepest descent path.}
\label{FigSmallS}
\end{center}
\end{figure}

We now deform the original contours in~\eqref{trkn} to the new contours of steepest-ascent and steepest-descent, which we call by the same names, $\Gamma_0$ and $\Gamma_1$.
We first deform the original contours to those in (a) of Figure~\ref{FigPrincValue} where $\Gamma_0$ is the contour of steepest-ascent and the part of $\Gamma_1$ except for the segment from $z_c^-$ to $z_c^+$ is the part of the contour of steepest-descent.
These contours can be divided as in (b) of Figure~\ref{FigPrincValue} and we have
\begin{equation}
\begin{aligned}\label{PVintegral}
	(\ref{trkn})&= \frac{-1}{(2\pi\I)^2} P.V. \oint_{\Gamma_0} \oint_{\Gamma_1}\frac{e^{M F_{u}(w)}}{e^{M F_{u}(z)}} \frac{dz\,dw}{(w-z)^2}
+\frac{1}{(2\pi\I)^2}\int_{\mathcal{C}}\oint_{\Gamma''_1}\frac{e^{M F_{u}(w)}}{e^{M F_{u}(z)}} \frac{dz\,dw}{(w-z)^2}.
\end{aligned}
\end{equation}
Here the first integral needs to be interpreted as the Principal Value due to the divergent terms in the integrand.
The second integral is from the contributions of the pole in the deformation of the contours.
The contours in the second double integral are defined as follows. The $w$-contour, $\mathcal{C}$, is a segment from $z_c^-$ to $z_c^+$  to the left of~$1$ and to the right of $0$.
The $z$-contour, $\Gamma''_1$, encircles the whole segment $\mathcal{C}$ but not $1$, see Figure~\ref{FigPrincValue}.
\begin{figure}[t]
\begin{center}
\psfrag{(a)}{(a)}
\psfrag{(b)}{(b)}
\psfrag{0}{$0$}
\psfrag{1}[c]{$1$}
\psfrag{=}[c]{$=$}
\psfrag{+}[c]{$-$}
\psfrag{zc}[c]{$z_c^+$}
\psfrag{zcBar}[c]{$z_c^-$}
\psfrag{G0}{$\Gamma_0$}
\psfrag{C}[c]{$\cal C$}
\psfrag{G1}{$\Gamma_1$}
\psfrag{G1I}{$\Gamma_1$}
\psfrag{G1II}{$\Gamma_1''$}
\includegraphics[width=\textwidth]{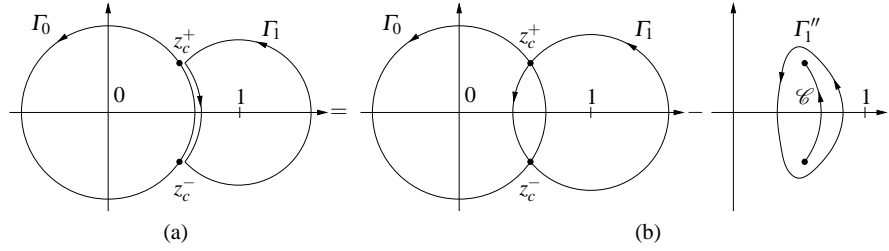}
\caption{The subdivision of the integration from (a) the ones in (\ref{trkn}) to (b) the ones in (\ref{PVintegral}).}
\label{FigPrincValue}
\end{center}
\end{figure}

Setting $Q(z):=\exp(M F_{u}(z))$, the Cauchy's integral formula implies that the second integral of~\eqref{PVintegral} equals
\begin{equation}
\frac{-1}{2\pi\I}\int_{\mathcal C}\frac{Q'(w)}{Q(w)} dw\,
=\frac{-M\left (F_{u}(z_c^+)-F_{u}(z_c^-) \right)}{2\pi\I}.
\end{equation}
Noting that $F_u(z_c^+)=\overline{F_u(z_c^-)}$, we have
\begin{equation}\label{contris32}
\frac{1}{(2\pi\I)^2}\int_{\cal C}\oint_{\Gamma''_1}\frac{e^{M F_{u}(w)-M F_{u}(z)}}{(w-z)^2}dz\,dw =\frac{-M \,\IM (F_{u}(z_c^+))}{\pi}.
\end{equation}
Observe that when $u'=(1+\sqrt{\gamma})^2$, the two critical points coincide and \mbox{$z_c:=z_c^{\pm}=\frac{1}{1+\sqrt \gamma}$}. In addition, $F_{M(1+\sqrt{\gamma})^2}(z_c)\in \R$. Thus
\begin{equation}\label{eq68}
\begin{aligned}
-\IM (F_{u}(z_c^+))&=\IM (F_{M(1+\sqrt \gamma)^2}(z_c)) - \IM (F_{u}(z_c^+)) \\
&=  \int_{u}^{M(1+\sqrt \gamma)^2} \IM\frac{d}{dv}F_v(z_c^+(v)) dv.
\end{aligned}
\end{equation}
Using the definition~\eqref{eq:Fud} of $F_u$ and the fact that $z_c^+(\nu)$ is a critical value, we find that $\frac{d}{dv}F_v(z_c^+(v))=\frac1{M} z_c^+(v)$.
From the formula (\ref{critpoints}) of $z_c^+$,
\begin{equation}\label{eq68-1}
\begin{aligned}
	\IM (z_c^+(v'M)) &= \frac{\sqrt{\nu'-(1-\sqrt{\gamma})^2}}{2\nu'}  ((1+\sqrt{\gamma})^2-\nu')^{1/2} \\
&\ge \frac{\epsilon}2 ((1+\sqrt{\gamma})^2-\nu')^{1/2}
\end{aligned}
\end{equation}
since $\nu'$ satisfies the condition~\eqref{eq:condup1}.
Therefore,~\eqref{eq68} implies that
\begin{equation}
	(\ref{contris32}) \geq  M \frac{\epsilon \pi}{3} ((1+\sqrt \gamma)^2-u')^{3/2}.
\end{equation}
Recall that \mbox{$\sqrt{\gamma}=(1-\rho)/\rho+O(t^{-1/3})$}, $M=\rho^2 t(1+O(t^{-1/3}))$, and $u'=u/M$ with \mbox{$u=t+s\chit$}. Then, we can choose a $s_0>0$ large enough (but fixed independently of $t$) such that for all $s\leq -s_0$ it holds $(1+\sqrt{\gamma})^2-u'\geq -c_1 s t^{-2/3}$ for some $c_1>0$.
Therefore for $u'$ satisfying~\eqref{eq:condup1}, there is a positive constant $c$ such that
\begin{equation}\label{eq:contrfrompole}
	\frac{1}{(2\pi\I)^2}\int_{\cal C}\oint_{\Gamma''_1}\frac{e^{M F_{u}(w)-M F_{u}(z)}}{(w-z)^2}dz\,dw
	\geq c (-s)^{3/2}
\end{equation}
uniformly in $t$.

We now show that the contribution of the Principal Value integral in (\ref{PVintegral}) is much smaller than~\eqref{eq:contrfrompole}. Indeed we will show that this is $O(1)$. This proves~\eqref{eq:Trestcs32} by taking the constant $c$ smaller than one in~\eqref{eq:contrfrompole}.

A direct computation shows that
\begin{equation}
	F_{u}''(z_c^+)=(1-\gamma)\frac{(z_c^+-\frac{1}{1+\sqrt \gamma})(z_c^+-\frac{1}{1-\sqrt \gamma})}{(z_c^+)^2(z_c^+-1)^2}.
\end{equation}
This implies
\begin{equation}
|F_{u}''(z_c^+)|\sim ((1+\sqrt \gamma)^2-u')^{1/2}\sim s^{1/2}t^{-1/3}
\end{equation} as $u'\to (1+\sqrt{\gamma})^2-s t^{-2/3}$, while for $(1-\sqrt{\gamma})^2+\epsilon\le u'<(1+\sqrt \gamma)^2-\epsilon$ we have $|F_{u}''(z_c^+)|=O(1)$. Thus, for the general $u'$ satisfying~\eqref{eq:condup1},
\mbox{$c_1 t^{-1/3} \le |F''_u(z_c^+)| \le c_2$} for some positive constants $c_1$ and $c_2$.
Hence \mbox{$O(t^{1/3})\le \sqrt{M F''_u(z_c^+)} \le O(t^{1/2})$}.
Let us choose the parts $V_0(z_c^\pm)$ and $V_1(z_c^\pm)$ of the paths $\Gamma_0$ and $\Gamma_1$ respectively whose size are $B (MF''_u(z_c^+))^{-1/2}$. Then both parts become smaller as $t\to \infty$.
Because the paths $\Gamma_0$ and $\Gamma_1$ are chosen to be steep descent, the contribution coming from \mbox{$\Gamma_0\times\Gamma_1\setminus \{V_0(z_c^+)\cup V_0(z_c^-)\}\times \{V_1(z_c^+)\cup V_1(z_c^-)\}$} is at most of order $O(1)$ if  $B$ is chosen large enough (but fixed).
Let us first consider the contributions from the intersecting contours $V_0(z_c^+)\times V_1(z_c^+)$ and $V_0(z_c^-)\times V_1(z_c^-)$.
Due to the symmetry, it is enough to consider the contribution of $V_0(z_c^+)\times V_1(z_c^+)$, given by
\begin{equation}
	I(z_c^+):=\frac{-1}{(2\pi\I)^2} P.V. \int_{V_0(z_c^+)} \int_{V_1(z_c^+)}e^{ M F_{u}(w)-M F_{u}(z)}\frac{1}{(w-z)^2}dz\, dw.
\end{equation}
Now we have to see if this integral is bounded by a constant. Since $z$ converges to $z_c^+$, we use the Taylor's series of $F_u$ in $z-z_c^+$. Since $z_c^+$ is a critical point, the function $F_u(z)$ in the exponent may be approximated as \mbox{$F_u(z_c^+)+\frac12 F_u''(z_c)(z-z_c^+)^2$}. It can be checked that the contributions from the higher order terms are negligible. Changing the variables as $z=z_c^+ +z'(M F''_u(z_c^+))^{-1/2}$, $w=z_c^+ +w' (M F''_u(z_c^+))^{-1/2}$,
we obtain
\begin{equation}
	I(z_c^+) \approx \frac{-1}{(2\pi\I)^2} P.V. \int_{\I \R} \int_{\R}
	\frac{e^{ \frac12 ({w'}^2-{z'}^2) } }{(w'-z')^2} dz'\, dw'\label{74}
\end{equation}
which is finite.

Let us now show that the contribution of the non-intersecting contours \mbox{$V_0(z_c^{\pm})\times V_1(z_c^{\mp})$} are also bounded from above by some constant. To that aim, $B$ being fixed, we assume that $s_0$ is chosen large enough so that $s_0\gg B$. This time the singularity term $1/|w-z|^2$ is bounded from above and one can easily deduce that for all $u'\in ((1-\sqrt \gamma)^2+\epsilon, (1+\sqrt \gamma)^2-s_0t^{-2/3})$,
$$\Big |\frac{-1}{(2\pi\I)^2} P.V. \int_{V_0(z_c^{\pm})} \int_{V_1(z_c^{\mp})}e^{ M F_{u}(w)-M F_{u}(z)}\frac{1}{(w-z)^2}dz\, dw\Big |\leq O(1).$$
Combining the whole, we have shown that the contributions from the first integral in (\ref{PVintegral}) is $O(1)$ and~\eqref{eq:Trestcs32} is proved for $u'\in [(1-\sqrt{\gamma})^2+\epsilon,(1+\sqrt{\gamma})^2-s_0 t^{-2/3})$.

We now consider the case where
\begin{equation}\label{eq:tempo2}
	u'\in (0,(1-\sqrt \gamma)^2+\epsilon).
\end{equation}
In~\eqref{eq:condup1}, we could have chosen $\epsilon>0$ small enough so that
\begin{equation}\label{eq:tempo}
	(1-\sqrt \gamma)^2+\epsilon< \frac12\big( (1+\sqrt \gamma)^2+ (1-\sqrt \gamma)^2\big).
\end{equation}
Consider the Laguerre Unitary Ensemble $\frac{1}{m-d}XX^*$ where $X$ is a \mbox{$(m-d)\times (m+d)$} random matrix with i.i.d.\ complex standard Gaussian entries.
Denote by \mbox{$\lambda_1\geq \lambda_2 \geq \cdots \geq \lambda_{m-d}$} its ordered eigenvalues.
By the definition of the correlation kernel $K_t$, we have
\begin{equation} \label{minotr1}
\Tr(K_t)= \EE \bigg( \sum_{i=1}^{m-d} \Id_I(\lambda_i)\bigg),
\end{equation}
where $I=(u', +\infty)$. This can be bounded below as
\begin{equation}
\EE \bigg( \sum_{i=1}^{m-d} \Id_I(\lambda_i)\bigg)\geq \EE \bigg( \sum_{i=1}^{m-d} \Id_{I_{\epsilon}}(\lambda_i)\bigg),\label{minotr2}
\end{equation}
where $I_{\epsilon}=((1-\sqrt \gamma)^2+\epsilon, +\infty).$
Now, we call on the results of~\cite{BMY03}, giving convergence rates for the spectral distribution of random sample covariance matrices. Let $\mathbf{F}_{m-d}$ denote the empirical probability distribution function associated to the spectral measure:
\begin{equation}
	\mathbf{F}_{m-d}(x)= \frac{1}{m-d}\sum_{i=1}^{m-d} \Id_{\lambda_i\le x}.
\end{equation}
Let also $\mathbf{F}$ be the cumulative distribution function of the Marchenko-Pastur distribution $\varrho$ defined by the density
\begin{equation}
\frac{d \varrho}{dx\,}=\frac{\sqrt{(u_+^c-x)(x-u_-^c)}}{2\pi x
}\Id_{[u_-^c, u_+^c]}(x),
\end{equation}
where $\gamma=\frac{m+d}{m-d}$ and $u_c^{\pm}=(1\pm \sqrt \gamma)^2$.
It is well known that $\mathbf{F}_{m-d}(x)\to \mathbf{F}(x)$ a.s. for all $x$. In~\cite{BMY03} it is proven that
\begin{equation}
\max_{x>0}|\EE(\mathbf{F}_{m-d}(x))-\mathbf{F}(x)|\leq (m-d)^{-1/2}.
\end{equation}
Then (\ref{minotr1}) and (\ref{minotr2}) imply that
\begin{equation}
	\Tr(K_t) \geq (m-d)(1-\mathbf{F}((1-\sqrt \gamma)^2+\epsilon))-(m-d)^{1/2}.
\end{equation}
With the condition~\eqref{eq:tempo} on $\epsilon$, $\mathbf{F}((1-\sqrt \gamma)^2+\epsilon)<1$ uniformly and since \mbox{$m-d=\rho^2t + O(t^{2/3})\to \infty$}, we find that
there exists a positive constant $C=C(\epsilon)$ such that
\begin{equation}\label{eq:trcase3}
	\Tr(K_t) \geq C (m-d)
\end{equation}
uniformly in $t$ and for $u'$ satisfying~\eqref{eq:tempo2}.
Now as $u=t-s\chit$ and $u\ge 0$, we have $(-s)^{3/2}= (t-u)^{3/2} (\chi/t)^{1/2}\le \chi^{1/2} t$.
Thus since $m-d=\rho^2t + O(t^{2/3})$,~\eqref{eq:trcase3} implies that
there exists a positive constant $c$ such that
\begin{equation}\label{eq:trcase4}
	\Tr(K_t) \geq c (-s)^{3/2}
\end{equation}
uniformly in $t$ and for $u'$ satisfying~\eqref{eq:tempo2}.
Thus~\eqref{eq:Trestcs32} is proved for $u'$ satisfying~\eqref{eq:tempo2}.
This completes the proof of Proposition~\ref{TheoBoundTrace}.
\end{proofOF}

\section{Proof of Corollary~\ref{corollario}}\label{sectCor}

Let us consider the rescaled height function
\begin{equation}
	H_t(w):=\frac{h_t(j(w))-[(1-2\chi)t +2w (1-2\rho) \chi^{1/3}t^{2/3}]}{-2\chi^{2/3}t^{1/3}},
\end{equation}
with $j(w)=(1-2\rho)t+2w\chi^{1/3}t^{2/3}$. By (\ref{eqFwt}), $F_w(s,t)=\Pb(H_t(w)\leq s)$. We have:
\begin{equation}
G_t(w):={\rm Var}(H_t(w))=\int_{\R} s^2 dF_w(s,t)-\left(\int_{\R} s dF_w(s,t)\right)^2,
\end{equation}
and, in the original variables,
\begin{equation}
{\rm Var}(h_t(j(w)))=(2\chi^{2/3}t^{1/3})^2 G_t(w).
\end{equation}
Using the notation $\delta:=(2\chi^{1/3}t^{2/3})^{-1}$, by (\ref{eqS})
\begin{equation}\label{eq80}
\begin{aligned}
\int_{\R}2\chi^{1/3}t^{2/3} S(j(w),t) f(w) dw&=\frac{\chi}{4}\int_{\R} \frac{G_t(w+\delta)-2 G_t(w)+ G_t(w-\delta)}{\delta^2} f(w)dw\\
&=\frac{\chi}{4}\int_{\R} G_t(w) \frac{f(w+\delta)-2 f(w)+ f(w-\delta)}{\delta^2}dw.
\end{aligned}
\end{equation}
By Theorem~\ref{Ourthm} and the fact that $\int_\R s\, dF_w(s)=0$ (see~\cite{BR00}), we have that $G_t(w)$ converges to $g_{\rm sc}(w)$ uniformly for $w$ in a compact set of $\R$. Therefore, for smooth test functions $f$ with compact support, as $t\to\infty$ this expression converges to
\begin{equation}
	\frac{\chi}{4}\int_{\R} g_{\rm sc}(w) f''(w) dw= \frac{\chi}{4}\int_{\R} g_{\rm sc}''(w) f(w)dw.
\end{equation}


\begin{thebibliography}{10}

\bibitem{BMY03}
Z.D. Bai, B.~Miao, and J.~Yao, \emph{Convergence rates of spectral
  distributions of large sample covariance matrices}, SIAM J. Matrix Anal.
  Appl. \textbf{25} (2003), 105--127.

\bibitem{BBP06}
J.~Baik, G.~{Ben Arous}, and S.~P\'ech\'e, \emph{Phase transition of the
  largest eigenvalue for non-null complex sample covariance matrices}, Ann.
  Probab. \textbf{33} (2006), 1643--1697.

\bibitem{BDJ99}
J.~Baik, P.~Deift, and K.~Johansson, \emph{On the distribution of the length of the longest increasing subsequence of random permutations}, J. Amer. Math. Soc. \textbf{12} (1999), 1119--1178.

\bibitem{BDMMZ01}
J.~Baik, P.~Deift, K.~McLaughlin, P.~Miller, and K.~Johansson, \emph{Optimal tail estimates for directed last passage site percolation with geometric random variables}, Adv. Theor. Math. Phys. \textbf{5} (2002), 1207--1250.

\bibitem{BFP09}
J.~Baik, P.L. Ferrari, and S.~P{\'e}ch{\'e}, \emph{{Limit process of stationary
  TASEP near the characteristic line}}, Comm. Pure Appl. Math. \textbf{63}
  (2010), 1017--1070.

\bibitem{BR00}
J.~Baik and E.M. Rains, \emph{Limiting distributions for a polynuclear growth
  model with external sources}, J. Stat. Phys. \textbf{100} (2000), 523--542.

\bibitem{BF08}
A.~Borodin and P.L. Ferrari, \emph{{Anisotropic growth of random surfaces in
  $2+1$ dimensions}}, arXiv:0804.3035 (2008).

\bibitem{BP08}
A.~Borodin and S. P{\'e}ch{\'e}, \emph{{Airy kernel with two sets of parameters in directed percolation and random matrix theory}}, J. Stat. Phys. \textbf{132} (2008), 275--290.

\bibitem{Fer90}
P.A. Ferrari, \emph{Shock fluctuations in asymmetric simple exclusion}, Probab.
  Theory Relat. Fields \textbf{91} (1992), 81--101.

\bibitem{FS05a}
P.L. Ferrari and H.~Spohn, \emph{Scaling limit for the space-time covariance of
  the stationary totally asymmetric simple exclusion process}, Comm. Math.
  Phys. \textbf{265} (2006), 1--44.

\bibitem{Jo00b}
K.~Johansson, \emph{Shape fluctuations and random matrices}, Comm. Math. Phys.
  \textbf{209} (2000), 437--476.

\bibitem{Lig76}
T.M. Liggett, \emph{Coupling the simple exclusion process}, Ann. Probab.
  \textbf{4} (1976), 339--356.

\bibitem{Ok02}
A. Okounkov, \emph{Infinite wedge and random partitions}, Selecta Math. (N.S.)
  \textbf{7} (2002), 57--81.

\bibitem{PS01}
M.~Pr{\"a}hofer and H.~Spohn, \emph{Current fluctuations for the totally
  asymmetric simple exclusion process}, In and out of equilibrium
  (V.~Sidoravicius, ed.), Progress in Probability, Birkh{\"a}user, 2002.

\bibitem{PS02b}
M.~Pr{\"a}hofer and H.~Spohn, \emph{Exact scaling function for one-dimensional
  stationary {KPZ} growth}, J. Stat. Phys. \textbf{115} (2004), 255--279.

\bibitem{Sp83}
H.~Spohn, \emph{Excess noise for a lattice gas model of a resistor}, Z. Phys. B
  \textbf{57} (1984), 255--261.

\bibitem{TW94}
C.A. Tracy and H.~Widom, \emph{{Level-spacing distributions and the Airy
  kernel}}, Comm. Math. Phys. \textbf{159} (1994), 151--174.

\bibitem{BKS85}
H.~van Beijeren, R.~Kutner, and H.~Spohn, \emph{Excess noise for driven
  diffusive systems}, Phys. Rev. Lett. \textbf{54} (1985), 2026--2029.

\bibitem{Wid02}
H.~Widom, \emph{On convergence of moments for random young tableaux and a
  random growth model}, Int. Math. Res. Not. \textbf{9} (2002), 455--464.

\end{thebibliography}
\end{document}